\documentstyle[12pt,epsfig]{article}
\newcommand{\mn}{{\mu\nu}}
\newcommand{\D}{{\Delta}}

\newcommand{\s}{{\sigma}}
\newcommand{\th}{{\theta}}
\newcommand{\p}{{\partial}}
\newcommand{\eps}{{\varepsilon}}

\newcommand{\be}{\begin{equation}}
\newcommand{\ee}{\end{equation}}
\newcommand{\bea}{\begin{eqnarray}}
\newcommand{\eea}{\end{eqnarray}}
\begin{document}
  \begin{flushright} \begin{small}
  DTP--MSU/00-14 \\ DF/IST-5.2000 \\gr-qc/0008076
  \end{small} \end{flushright}
\vskip1cm

\begin{center}
{\Huge\bf No-go theorem for false vacuum }

\smallskip
{\Huge \bf black holes}

\vskip1cm
 {\bf Dmitri V. Gal'tsov}\footnote{Supported by RFBR; Email: galtsov@grg.phys.msu.su}

\smallskip
{\it Department of Theoretical Physics,}   \\
       {\it  Moscow State University, 119899, Moscow, Russia}

\smallskip
and\\ {\bf Jos\'e P. S. Lemos}\footnote{E-mail: lemos@kelvin.ist.utl.pt }

\smallskip
  { CENTRA, Departamento de F\'{\i}sica,
              Instituto Superior T\'ecnico,} \\
  {\it Av. Rovisco Pais 1, 1096 Lisboa, Portugal}
\end{center}
\date{\today}
\vskip1cm
\begin{abstract}
We study the possibility of non-singular black hole solutions in the
theory of general relativity coupled to a non-linear scalar field with
a positive potential possessing two minima: a `false vacuum' with
positive energy and a `true vacuum' with zero energy. Assuming that
the scalar field starts at the false vacuum at the origin and comes to
the true vacuum at spatial infinity, we prove a no-go theorem by
extending a no-hair theorem to the black hole interior: no smooth
solutions exist which interpolate between the local de Sitter solution
near the origin and the asymptotic Schwarzschild solution through a
regular event horizon or several horizons.
\end{abstract}
\newpage

\section{Introduction}
In a recent paper Daghigh, Kapusta and Hosotani \cite{DaKaHo00}
proposed a new type of a black hole in the theory of general
relativity coupled to a non-linear scalar field with a
potential. A quartic non-symmetric potential was assumed to
have one minimum of positive energy (false vacuum) and another
of zero energy (true vacuum). In the false vacuum state, a
possible solution is the de Sitter metric, while for the scalar
field at the true vacuum one can assume the Schwarzschild
metric. The `false vacuum black hole' was suggested to be given
by the Schwarzschild solution outside the event horizon and the
de Sitter solution inside the black hole. The scalar field was
supposed to have a constant false vacuum value inside the black
hole and a constant true vacuum value outside with a finite jump
at the horizon. Direct matching of de Sitter spacetime with the
Schwarzschild solution on the horizon had been suggested earlier
\cite{Go81,ShZh88} and shown to be incorrect
\cite{Gr85,PoIs88,GrSo89}. However, it was argued in
\cite{DaKaHo00} that the matching of these solutions can be
achieved within a more general parameterization of the static
metric by two different functions due to the jump of the product
$g_{tt}g_{rr}$.  If correct, this proposal could lead to
intriguing physical implications like new kinds of black hole
remnants or the possibility that we live inside an enormous
black hole.

Meanwhile, the problem of incorporating the de Sitter metric inside
the black hole is by no means new. The idea that sigularitities can be
avoided by matter with an `inflationary' equation of state was
forwarded by Gliner \cite{Gl66} long before the inflation scenario was
proposed in cosmology.  It was also suggested that the limiting
curvature principle of Markov \cite{Ma84} may be realized in the black
hole context as the appearence of a de Sitter world instead of the
singularity \cite{FrMaMu90,BaFr96}.  A similar issue was discussed
using the idea of vacuum polarisation \cite{PoIs88} and in the
`cutoff' curvature approach \cite{Po89,Mo91}.  Some phenomenological
matter sources were suggested ensuring a de Sitter nucleus inside the
black hole \cite{Dy92,Dy99}.  It was also shown that, under
perturbations, the de Sitter metric emerges inside a charged black
hole in the development of an instability of the internal Cauchy
horizon, a phenomenon called mass-inflation with an exponential growth
of the local mass function \cite{PoIs90}. Such a growth has also been
observed inside static Einstein-Yang-Mills black holes when the
singularity is approached \cite{DoGaZo96,GaDoZo97,BrLaMa98}.  A
closely related subject -- the avoidance of singularities inside black
holes not neccessarily related to the de Sitter solution -- has
attracted much attention recently. Such a possibility was described by
Bardeen in 1968 \cite{Ba68} as a modification of the
Reissner-Nordstr\"om metric. Recently a non-linear electrodynamics
lagrangian was found producing a Bardeen type metric
\cite{AyGa99}. Phenomenological sources for non-singular black holes
were discussed in \cite{MaMaSe96,Bo97}. In addition, we mention
some investigations on the dynamics of time-dependent bubbles with
false vacuum inside (and hence de Sitter metric inside) and a black
hole metric outside (for a recent discussion see \cite{AlLoTr99}).
Thus, the main surprise of the paper \cite{DaKaHo00} is a claim that
an internal de Sitter metric can be accommodated to a black hole as a
static solution within such a simple model as the scalar theory with a
non-symmetric potential.

Here we investigate the problem of matching the de Sitter and the
Schwarzschild metrics in the context of this model in more detail.
First we show that the piecewise solution suggested in \cite{DaKaHo00}
cannot be interpreted in terms of distributions and is likely to
demand additional singular matter sources at the horizon. Then we
discuss the possibility of more complicated smooth solutions assuming
more general positive non-symmetric potential with two minima. First,
using only a local analysis we explore the possibility of matching the
internal and external solutions at the horizon attached to the local
maximum of the potential. It turns out that both inside and outside
the black hole one encounters timelike regions (with $\p_t$
spacelike). Finally, we discuss the problem from a global viewpoint,
in the spirit of no-hair theorems. We prove a no-go theorem by
extending the no-hair argument to the black hole interior and showing
that the model does not admit smooth solutions in which the scalar
field starts at the false vacuum at the origin and comes to the true
vacuum at infinity with one or several horizons at finite values of
the radii of spherical sections. Some open possibilities for false
vacuum black holes are then briefly discussed.
\section{The model}
Consider general relativity coupled to a real scalar field theory
\be  \label{A}
S=\frac{1}{4\pi}\int\left(-\frac{R}{4}+
\frac12 g^{\mn}\p_\mu\phi\p_\nu\phi-V(\phi)\right)\sqrt{-g}\,d^4x
\ee
with a smooth potential $V$ for which we assume a general behavior as shown
in Fig.1.
In what follows the potential is not necessarily quartic. We need only
that near minima, $-\phi_1$ and $\phi_2$, it can be approximated by parabolae
\bea
V&=& V_1+\frac12 k_1 (\phi+\phi_1)^2 +O((\phi+\phi_1)^3),\label{V1}\\
V&=& \frac12 k_2 (\phi-\phi_2)^2 +O((\phi-\phi_2)^3),\label{V2}
\eea
with positive constants $V_1, k_1, k_2$,
and the local maximum is at $\phi=0$, with $V(0)=V_0>0$. We assume that
the potential goes to infinity as a finite power of $|\phi|$
as $\phi\to\pm\infty$, and that the derivative
\be
V_{\phi}=\frac{dV}{d\phi}.
\ee
is finite for finite $\phi$.

Assuming spherical symmetry and staticity we write the metric in the
curvature gauge
\be \label{ds}
ds^2=\s^2 Ndt^2-\frac{dr^2}{N}-r^2d\Omega \, ,
\ee
where $\s$ and $N$ are functions of $r$. For $N$ we will also
use the following two parametrizations:
\be
N=1-\frac{2m}{r}=\frac{\D}{r} \, ,
\ee
where $\D=r-2m$.
The equations of motion following from the action (\ref{A}) read
\bea
\frac{\s'}{\s}&=&r\phi'^2,\label{seq}\\
m'&=&r^2\left(\frac12 N\phi'^2+V\right),\label{meq}\\
\left(r^2 N\s \phi'\right)'&=&\s r^2V_{\phi}\label{fseq}.
\eea
The variable $\s$ can be excluded from the Eq.~(\ref{fseq}) using Eq.
(\ref{seq}).  Denoting $\xi=r\phi'$, one can rewrite the equations of
motion as a system of three first order equations for $\phi$, $\xi$,
$\D$
\bea
\D'&=&-\frac{\D}{r}\xi^2 +U,\label{deq}\\
\D\xi'&=&r^2 V_\phi-\xi U,\label{xeq}\\
\phi'&=&\frac{\xi}{r},\label{feq}
\eea
where
\be \label{U}
U=1-2r^2V,
\ee
and an equation for $\s$, i.e. Eq.  (\ref{seq}), which can be solved
once the solution of the system (\ref{deq}-\ref{feq}) is found.
\begin{figure}[t]
\centerline{\epsffile{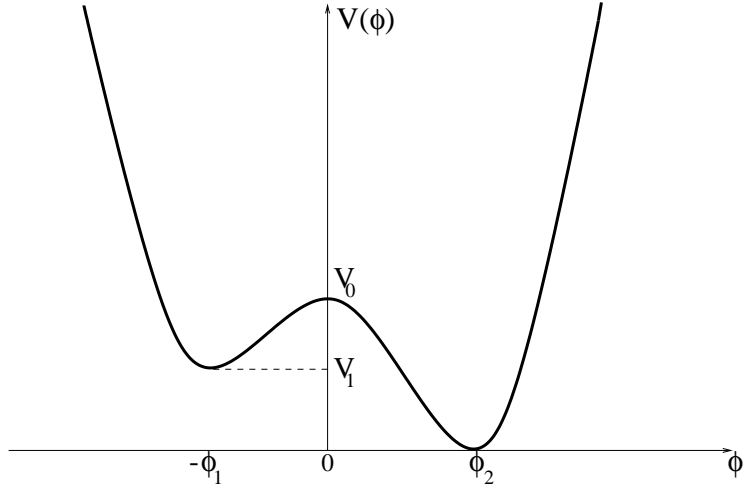}} \vskip0.5cm
\caption{  The potential $V$ as a function of $\phi$.}
\label{fig1}
\end{figure}
A mechanical analogy is useful for the scalar field equation (\ref{fseq})
regarding $r$ as the mechanical time. Then the role of the potential will be
played by $-V$ in the spacelike region ($\D>0$) and by $V$ in the
timelike region ($\D<0$, if any). Thus, a `particle' departing from
one of the vacua will be forced to climb the barrier if $\D>0$, but
will be turned back if $\D<0$.

One trivial solution to the system (\ref{deq}-\ref{feq}) corresponds to
the scalar field sitting at the left local minimum of the potential
(false vacuum):
\be
\phi\equiv -\phi_1,\quad \s\equiv 1,\quad N=1-\frac{r^2}{r_c^2}.
\ee
This is the de Sitter  spacetime with the cosmological horizon   at
\be \label{rc}
r=r_c=\sqrt{\frac{3}{2V_1}}.
\ee
Another trivial solution is the  Schwarzschild metric. This corresponds
to the true vacuum
\be
\phi\equiv\phi_2,\quad \s\equiv 1,\quad N=1-\frac{M}{r},
\ee
with finite mass $M$. Note that any constant value of the metric function
$\s$ may be rescaled to unity by a time scaling.
\section{Piecewise solution}
Let us try, following \cite{DaKaHo00}, to construct a piecewise
solution to the Eqs. (\ref{deq}-\ref{feq}) assuming for the
scalar field a step-like behaviour \be \label{fiteta}
\phi=-\phi_1 \th(r_h-r) + \phi_2 \th(r-r_h), \ee i.e., the
false vacuum value inside the black hole and the true vacuum
value outside. Similar matching within the parameterization of
the metric by a single function of the radial variable was
discussed earlier in \cite{Gr85,GrSo89} and shown to lead to
instabilities. Our current model differs in that the metric
functions $g_{tt}$ and $g_{rr}$ are independent, so these
results do not apply directly.

The field equations will be satisfied for $r<r_h$ with the de
Sitter metric \be \s=\s_{0},\quad m=\frac{r^3}{2r_h^2}, \ee
(where the constant value $\s_0$ can be removed by a rescaling
of time), and for $r>r_h$ with the Schwarzschild metric \be
\s=\s_{\infty},\quad m=\frac{r_h}{2}, \ee the event horizon
coinciding with the de Sitter cosmological horizon. This
suggests the following representation for the metric functions:
\bea
\s&=&\s_{0}\th(r_h-r)+\s_{\infty}\th(r-r_h),\label{steta}\\
m&=&\frac{r^3}{2r_h^2}\th(r_h-r)+ \frac{r_h}{2}\th(r-r_h).
\label{mteta} \eea However, an attempt to check whether this
solution remains true at the horizon $r=r_h$ in the sense of
distributions fails because the system (\ref{deq}-\ref{feq}) is
highly non-linear. This could be expected: distribution-valued
metrics and curvature can be introduced in general relativity
only under special conditions (see, e.g., \cite{Ga99}). Perhaps
more satisfactory results could be obtained in the framework of
the generalized functions approach by Colombeau \cite{Co84},
but it will be enough for our purposes here to use a simple
regularization for the delta-function \be \label{dreg}
\delta(x)=\lim_{\eps\to 0}\delta_\eps(x),\quad
\delta_\eps(x)=\frac{\eps}{\pi(x^2+\eps^2)}. \ee
Differentiation of Eq.~(\ref{fiteta}) gives \be
\xi=r_h(\phi_1+\phi_2)\delta(x), \quad x=r-r_h, \ee while for
the derivative of $m$ we obtain \be m'=\frac{3(1+x)^2}{2}
\th(-x). \ee Near $x=0$ one also has \be \label{Dx}
\D=x\left[\th(x)-2\th(-x)\right]\, , \ee therefore at the right
hand side of the Eq.(\ref{meq}) one encounters a product
$x\delta^2(x)$. Its regularized version \be
x\delta_\eps(x)^2=\frac{\eps^2}{\pi^2}\frac{x}{(x^2+\eps^2)^2}
\ee vanishes at $x=0$. Moreover, it can be checked that for any
`good' function $\varphi(x)$ one has: \be \lim_{\eps\to 0} \int
\varphi(x) x\delta_\eps^2(x) dx =0. \ee Dropping therefore the
first term in the right hand side of Eq.~(\ref{meq}) one can
verify the validity of this equation at $x=0$ in view of the
formula \be V=\frac{3}{2r_h^2}\th(-x). \ee

However, the situation with other equations is less satisfactory. For
the solution under consideration $V_\phi=0$, while the left hand side
of the Eq.~(\ref{fseq}) contains a delta-singularity. Then
Eq.~(\ref{seq}) shows that the jump of $\sigma$ at $x=0$ is infinite:
\be
\ln\s\Bigg|_{r_h-0}^{r_h+0}=r_h(\phi_1+\phi_2)^2 \delta(0),
\ee
(this is easy to check using the regularization (\ref{dreg})).  The
structure of other singularities in the Einstein tensor is likely to
exhibit the presence of extra matter at the horizon surface.  The
Einstein tensor for the metric (\ref{ds}) reads:
\bea
G_t^t&=& \frac{2m'}{r^2},\\
G_r^r&=& \frac{2}{r^2}\left(m'-\Delta\frac{\s'}{\s}\right),\\
G_\th^\th&=& -\frac{1}{2r}\left( \Delta''+
2\D \frac{\s''}{\s}+\left(3\D'-\frac{\D}{r}\right)\frac{\s'}{\s}\right),
\eea
while the right hand side of the Einstein's equations for the scalar field is
given by
\bea
8\pi T_t^t&=& N\phi'^2+2V,\label{Ttt}\\
8\pi T_r^r&=&-N\phi'^2+2V,\label{Trr}\\
8\pi T_\th^\th&=&N\phi'^2+2V.\label{Tthth}
\eea
The $tt$-equation coincides with the Eq.(\ref{meq}) and it is satisfied
at the horizon. The $rr$-equation is satisfied as well after omitting
the products $x\delta^2(x)$ as we argued above. But the situation is
more complicated with the $\th\th$-equation. Indeed, differentiating
(\ref{Dx}) one obtains a delta-term
\be
\D''=3\delta(r-r_h),
\ee
which is not canceled by other terms in this equation. The same is
true for the singular term containing $\D'\s'/\s$.  All this looks
like showing the presence of extra matter at the horizon.  This is not
very surprising: the piecewise model emerges when the scalar theory is
treated like a vacuum theory with a step-like cosmological
constant. But if the cosmological constaint is variable, the Bianchi
identities require additional matter to be invoked \cite{OvCo98}.

Therefore it is hard to adopt the piecewise metric (\ref{steta}),
(\ref{mteta}) as a true solution to the Einstein equations with a
scalar source.
\section{An attempt of a smooth matching at $\phi=0$}
One can imagine that a true solution with similar properies exists
which is deformed from the above simple form in the vicinity of the
horizon.  Here we perform a purely local analysis of the behavior of
the presumed smooth solution near the horizon assuming that the latter
corresponds to a `natural' point $\phi=0$, i.e., to the local maximum
of the potential shown in Fig. 1. Then $V_\phi(r_h)=0$ and we get from
the Eqs. (\ref{deq},\ref{xeq})
\be
\D'_h=U_h,\quad \xi_h U_h=0.
\ee
(Here we also used the fact that $\D\xi^2$ vanishes at the horizon,
this remains true even if $\xi$ diverges, see the next section). So either
\[
i) \quad U_h=0, \quad \xi_h \;\mbox{arbitrary},
\]
or
\[
ii)\quad \xi_h=0, \quad U_h \;\mbox{arbitrary}.
\]
Consider first the case i). Then $\D_h'=0$ and the horizon is degenerate
($\D$ has a zero of the second order). Therefore near the horizon
\be \label{degener}
\D=\frac{\alpha}{2} x^2 +O(x^3),
\ee
where $x=r-r_h$, and $\alpha$ should be positive to ensure the timelike
character of the Killing vector $\p_t$ outside the horizon.  Now expand the
Eq.~(\ref{deq}) to linear order in $x$.  Equating the linear terms one
gets:
\be
\alpha=U'_h,
\ee
with $U'_h$ being the value of the derivative of (\ref{U}) at the horizon.
It can be represented as
\be
U'_h=-4r_hV_h-2r^2_h V_\phi\Big|_{r_h}\phi'_h,
\ee
where, by the assumption that the horizon is at the maximum of the potential,
\be
V_\phi\Big|_{r_h}=0,
\ee
and by the assumption i)
\be
2r_h^2V_h=1.
\ee
It follows that
\be \label{alfa}
\alpha=-\frac{2}{r_h},
\ee
therefore $\p_t$ is spacelike outside the horizon.
By definition,
the event horizon is the largest root of the equation $\D=0$, so this
solution can not describe a black hole.

Consider now the case ii). If $\xi_h=0$ and $U_h\neq 0$ an expansion
for $\D$ will contain a linear term
\be
\D=U_h x+\frac{\alpha}{2} x^2 +O(x^3),
\ee
while $\xi$ starts as
\be
\xi=\xi'_hx+O(x^2).
\ee
Now, collecting linear terms in the Eq.~(\ref{deq}) we get again
Eq.~(\ref{alfa}) for $\alpha$, so the expansion of $\D$ will read
\be
\D=-\frac{1}{r_h}\left(x-\frac12 U_h r_h\right)^2 + \frac 14 U_h^2 r_h
+O(x^3)\, .
\ee
Now the region with the timelike $\p_t$ is between the horizons,
while outside it $\p_t$ is spacelike again.

So the result of our attempt is disapointing: one cannot
smoothly match the solutions with both interior and exterior
spacelike metrics at the event horizon attached to the `natural'
point $\phi=0$.

It is worth noting that the matching of de Sitter and
Reissner-Nordstr\"om metrics {\em is} possible along the
interior Cauchy horizon for some special choice of parameters
\cite{BaIs91}.
\section{No-go theorem}
Now we would like to study the problem in a more general setting:
whether a static spherically symmetric smooth solution to the model (\ref{A})
exists which is locally de Sitter near the origin
(with $\p_t$ timelike) and asymptotically Schwarzschild with one
or several regular horizons in the intermediate region.
More precisely, we will assume that  near the origin the space-time is flat:
\be \label{Ns0}
\quad N(0)=1,\quad \s(0)=\s_0\neq 0,
\ee
(the constant value $\s_0$ can be eliminated locally by rescaling of time),
and the scalar field is at the false vacuum
\be
\quad\phi(0)=-\phi_1;
\ee
while at infinity the solution is asymptotically flat
\be \label{Ns}
N=1-\frac{2M}{r}+O\left(\frac{1}{r^2}\right), \quad \s(\infty)=1,
\ee
and the scalar field is at the true vacuum:
\be
\phi(\infty)=\phi_2.\;\;
\ee

 We will give a non-existence proof in two steps: first we invoke the no-hair
argument for the exterior region, and then we extend it to the black hole
interior. The   exterior no-hair theorem for an Abelian Higgs model
was proven by Adler and Pearson \cite{AdPe78} (for an improved version
see \cite{Gi91,La93}). In \cite{AdPe78} the Goldstone model was also
considered for the Mexican hat potential. Here we deal with the real scalar
field with a more general positive potential, so it is worth giving
the no-hair proof explicitly.

The finiteness of the ADM mass $M$ imposes restrictions on the first
subleading term in the asymptotic expansion of the scalar field at infinity.
Integrating the Eq.~(\ref{meq}) from the event horizon (the maximal root
of $\D$) to infinity we find for the ADM mass
\be \label{intM}
  M= \frac{r_h}{2}+\int_{r_h}^{\infty}
\left(\frac12 N\xi^2+r^2V\right)dr,
\ee
where the first term is the `bare' mass, and the second one is the
contribution of the scalar hair. Both terms in the integrand are positive
semidefinite, so each of them should be integrable.
For convergence of the second term
at infinity it is necessary that $V$ as a function of $r$ decays
faster than $r^{-3}$.  In view of the Eq.~(\ref{V2}) this translates
to
\be\label{3/2}
\phi=\phi_2+o\left(\frac{1}{r^{3/2}}\right),
\ee
as $r\to\infty$. This implies
\be \label{xi3/2}
\xi=o\left(\frac{1}{r^{3/2}}\right),
\ee
so the first term in Eq.~(\ref{intM}) is also integrable at infinity.
 From the Eq.~(\ref{seq}) it then follows that at infinity
\be
\s=1-o\left(\frac{1}{r^3}\right),
\ee
with a negative subleading term, where the constant was set to unity
in conformity with the assumption (\ref{Ns}). With this normalization,
an integration
of the Eq.~(\ref{seq}) gives
\be \label{sexp}
\s=\exp\left(-\int_r^\infty \xi^2\frac{dr}{r}\right),
\ee
so $\s$ is bounded to the interval
\be
0< \s \leq 1.
\ee
Here the lower bound $\s=0$ could be reached at a point where the integral
in the exponential diverges. However in view of the assumption (\ref{Ns0})
$\s$ remains non-zero at the origin, and being a non-decreasing function,
remains strictly positive elsewhere.

We now discuss the behaviour of the solution near the event horizon.
An assumption of regularity implies that the mixed components of
the energy-momentum tensor (\ref{Ttt}-\ref{Tthth}) are finite, so
$\xi$ should satisfy
\be \label{NxiC}
\lim_{r\to r_h}N\xi^2 =C^2,
\ee
with some $C\geq 0$, while $\phi$ should not diverge in view of our
assumptions about the potential $V$ (also implying the finiteness of the
derivative $V_\phi$). These conditions ensure the
convergence of the integral (\ref{intM}) at the horizon.
One can obtain a stronger condition on $C$ using the field equations.
 From the Eqs.~(\ref{deq},\ref{xeq}) one derives
\be
(rN\xi)'+\frac{N}{r}\xi^3=r^2V_\phi.
\ee
Substituting here $\xi$ from the Eq.~(\ref{NxiC}) we obtain in the
right vicinity of the horizon
\be
\frac{CN'r_h}{2\sqrt{N}}+\frac{C^3}{\sqrt{N}}+\mbox{finite terms}=r^2 V_\phi.
\ee
The right-hand side of this equation remains finite as $r\to r_h$, while
the first two terms on the left-hand side are positive semidefinite
(recall that $N'_h\geq 0$ ) and diverge
(for a degenerate horizon only the second term diverges).
Therefore $C=0$, i.e. we get a stronger condition on $\xi$:
\be \label{Nxi2}
\lim_{r\to r_h}N\xi^2 =0.
\ee
(An alternative proof of this relation follows from the convergence
of the integral in the exponential in the Eq.~(\ref{sexp}).) From
the Eq.~(\ref{NxiC}), a weaker condition also holds
\be \label{Nxi1}
\lim_{r\to r_h}N\xi =0.
\ee

Now we can give a no-hair proof similar to that of Adler and Pearson
\cite{AdPe78}.From the Eqs. (\ref{fseq}), (\ref{deq}) and (\ref{seq})
one derives the following identity:
\be
 \frac{d}{dr}\left[\s \left(2r^2V-\frac{\D\xi^2}{r}\right)\right]=
 \s\left[\frac{\xi^2}{r}\left(1+\frac{\D}{r}\right)+4rV\right] \, .
\ee
Integrating it from the horizon to infinity and taking into account
the condition (\ref{Nxi2}), boundedness of $\s$, and the relations valid
at $r\to\infty$
\bea
r^2 V&=&o\left(\frac{1}{r}\right), \\
\frac{\xi^2 \D}{r}&=&o\left(\frac{1}{r^3}\right),
\eea
which follow from the Eqs. (\ref{V2}), (\ref{3/2}) and (\ref{xi3/2}),
we obtain
\be
\int_{r_h}^\infty \s\left[\frac{\xi^2}{r}\left(1+\frac{\D}{r}\right)
+4rV\right]dr=-2r_h^2V_h\s_h ,
\ee
where $V_h$ is the value of the potential at the event horizon.
Since   $\s$ is strictly positive, the potential is positive semidefinite,
and $\D\geq 0$ everywhere
in the integration region, it follows that both the left hand side and
the right hand side of this equation are strictly zero, which implies
\be
\phi\equiv\phi_2
\ee
for all $r\geq r_h$. Thus the event horizon should correspond to the
absolute minimum (true vacuum) of the potential. Consequently, the black hole
is exactly Schwarzschildean for an external observer.

Now let us look to the interior of the black hole.  Since we assume
that the Killing vector $\p_t$ is timelike in the vicinity of the
origin, there are two possibilities: either there exist an internal
Cauchy horizon at some $r=r_-$ (more generally an odd number of
internal horizons), or the event horizon at $r=r_h$ is degenerate. The
second possibility is a limiting case of the first, so we assume that
there are two solutions $r=r_\pm$ of the equation $\D=0$ such that
$r_-\leq r_+=r_h$ (the generalization of the following proof to a
finite odd number of internal horizons is straightforward). Generalizing
the condition (\ref{Nxi1}) we find
\be
\lim_{r\to r_{\pm}} N\phi'=0.
\ee
At the origin $\phi$ has a finite limit $-\phi_1$
and is (by assumption) a smooth function, therefore
\be
\lim_{r\to 0} r^2\phi'=0.
\ee
Then, integrating the Eq.~(\ref{fseq}) form the origin to $r_-$, and taking
into account boundedness of  $\s$  we get the relation
\be
r^2\s N\phi'\Bigg|_0^{r_-}=0=\int_0^{r_-}\s r^2 V_\phi dr.
\ee
It follows that either
\[\qquad i)\quad \;\phi(r)=-\phi_1 \quad \mbox{for all}\;\; r\in [0, r_-], \]
i.e. the solution is exactly de Sitter up to $r=r_-$ in which case
$r_-=r_c$ given by the Eq.~(\ref{rc}), or
\[ii)\quad \phi(r_-) \in (0, \phi_2], \;\qquad\qquad\]
i.e. the inner horizon is attached to the right wing $(0, \phi_2]$ of the
potential curve, see  Fig.1.

The case i) means that there is a global solution such that $\phi$ is
identically constant in the finite interval $[0, r_-]$, $\phi$ is
equal to a different constant along the semi-axis $[r_+, \infty)$, but
$\phi$ is varying in the interval $(r_-, r_+)$.  Impossibility of a
smooth solution of such a kind can be made clear from a mechanical
analogy: the particle sitting at the bottom of the potential well for
a finite time cannot start moving unless it is pushed, since all the
derivatives of $\phi$ at the initial moment are zero by continuity. To
get an idea how this follows from the field equations, consider a
vicinity of the event horizon $r=r_+$, where $V=0$ and hence $U=1$,
while $\D=r-r_+=x$. Then the leading terms in the Eq.~(\ref{xeq}) give
the following equation for $\xi$:
\be
x\xi'=-\xi.
\ee
The singular solution $\xi=1/x$ for $x<0$ can not be matched to
$\xi\equiv 0$ for $x>0$ (and does not satisfy (\ref{Nxi2})),
therefore the correct solution is $\xi=0$.

Now consider the case ii). Integrating the
Eq.~(\ref{fseq}) from  $r_-$ to $r_+$ we obtain
\be
r^2\s N\phi'\Bigg|_{r_-}^{r_+}=0=\int_{r_-}^{r_+}\s r^2 V_\phi dr.
\ee
Here $V_\phi\leq 0$, and therefore $V_\phi\equiv 0$ in
$[r_-,r_+]$ which implies
\be
r_-=r_+, \quad \phi(r_-)=\phi_2,
\ee
that is the event horizon is degenerate.  But then we come back to the
step like solution (\ref{fiteta}) which faces the problems discussed
in the Sec.~2 and should be ruled out by the smoothness
assumption. This completes the proof.
\section{Conclusion}
Our results are the following. First, we have shown that the piecewise
false vacuum black hole  presented in \cite{DaKaHo00} can not be
interpreted in terms of distributions and apparently is not a solution
to the Einstein-scalar field equations without additional matter sources.
Second, we have extended the no-hair argument to the black hole interior
and have shown that there are no smooth solutions to the scalar model
with a non-symmetric potential which interpolate
from the false vacuum inside (in the region with the timelike Killing
vector $\p_t$) and the true vacuum outside through the horizon(s).

One could ask whether it is possible to weaken some of the assumptions
made in order to reopen the possibility of such or similar
configurations. The first assumption is the positivity and the shape
of the potential. In fact, for static spherically symmetric
configurations one can invert the roles of the scalar field $\phi$ and
the potential $V(\phi)$: one chooses the desired behavior of $\phi$
determining afterwards the potential through the equations. In such a
way some potentials were found for which scalar hair does exist
\cite{BeLe95,DeLe96}. The main problem here is, of course, whether
these potentials are physically reasonable.  The second assumption is
asymptotic flatness.  It was found \cite{ToMaNa99} that in an
asymptotically de Sitter spacetime scalar hair does exist within a
similar model. Finally one could consider a non-minimal coupling where
no-hair theorems also get modified \cite{MaBe96,Be96}.

\vskip 1cm
\noindent {\large {\bf Acknowledgements}}

\bigskip
The authors thank anonymous referees for indicating the
references \cite{Go81,ShZh88,Gr85,GrSo89,BaIs91} and useful
comments. D.G. is grateful to CENTRA/Instituto Superioro
T\'ecnico (Lisbon) and GTAE for hospitality and support in
August 2000 when this paper was written.  His work was also
partially supported by the Russian Foundation for Basic
Research under the grant 00-02-16306. JPSL acknowledges a grant
from FCT through the project ESO/PRO/1250/98, and thanks Observat\'orio
Nacional - Rio de Janeiro for hospitality.


\newpage
\begin{thebibliography}{99}

\bibitem{DaKaHo00}
R.~G. Daghigh, J.~I. Kapusta and Y.~Hosotani,
False vacuum black holes and Universes, 2000, gr-qc/0008006.

\bibitem{Go81}
P.~F.~Gonzales-Diaz,
Lett. Nuovo Cim. {\bf 32} (1981) 161.

\bibitem{ShZh88}
W.~Shen and S.~Zhu,
Phys. Lett. {\bf A126} (1988) 229.

\bibitem{Gr85}
O.~Gron,
Lett. Nuovo Cim. {\bf 44} (1985) 177.

\bibitem{PoIs88}
E.~Poisson and W.~Israel, Class. Quant. Grav. {\bf 5} (1988)
L201.

\bibitem{GrSo89}
O.~Gron and H.~Soleng,
Phys. Lett. {\bf A138} (1989) 89.

\bibitem{Gl66}
E.~Gliner,
Sov. Phys. JETP, {\bf 22} (1966) 378;
see also gr-qc/9808042.

\bibitem{Ma84}
M.~A. Markov,
Ann. Phys. (NY) {\bf 155} (1984) 333.

\bibitem{FrMaMu90}
V.~P. Frolov, M.~A. Markov, and V.~F. Mukhanov,
Phys. Lett. {\bf B216} (1989) 272;
Phys. Rev. {\bf D41} (1990) 3831.

\bibitem{BaFr96}
C.~Barrabes and V.~P.~Frolov,
Phys. Rev. {\bf D53} (1996) 3215.



\bibitem{Po89}
J.~Polchinski,
Nucl. Phys. {\bf B 325} (1989) 619.

\bibitem{Mo91}
D.~Morgan,
Phys. Rev. {\bf D43} (1991) 3144.

\bibitem{Dy92}
I.~G. Dymnikova,
Gen. Rel. Grav. {\bf 24} (1992) 235.

\bibitem{Dy99}
I.~G. Dymnikova,
Phys. Lett. {\bf B472} (2000) 33.

\bibitem{PoIs90}
E.~Poisson and W.~Israel,
Phys. Rev. {\bf D41} (1990) 1796.

\bibitem{DoGaZo96}
E.E. Donets, D.V. Gal'tsov, and M.Yu. Zotov,
Phys. Rev. {\bf D 56} (1997) 3459.

\bibitem{GaDoZo97}
D.V. Gal'tsov, E.E. Donets,  and M.Yu. Zotov,
JETP Lett. {\bf 65} (1997) 895.

\bibitem{BrLaMa98}
P. Breitenlohner, G. Lavrelashvili, and D. Maison,
Nucl. Phys. {\bf B 524} (1998) 427.

\bibitem{Ba68}
J.~Bardeen, in Proceedings of GR5, Tbilisi, URSS, 1968.

\bibitem{AyGa99}
E.~Ay\'on-Beato and A.~Garcia,
Phys. Rev. Lett. {\bf 80} (1998) 5056.

\bibitem{MaMaSe96}
M.~Mars, M.~M. Martin-Prats and J.~M.~M. Senovilla,
Class. Quant. Grav. {\bf 13} (1996) L51.

\bibitem{Bo97}
A.~Borde,
Phys. Rev. {\bf D55} (1997) 7615.

\bibitem{AlLoTr99}
G.~L. Alberghi, D.~A. Lowe and M.~Trodden, 
JHEP {\bf 9907} (1999) 020.



\bibitem{Ga99}
D.~Garfinkle,
Class. Quant. Grav. {\bf16} (1999) 4101.

\bibitem{Co84}
J.~F. Colombeau,
{\em New Generalized Functions and Multiplication of Distributions
(North-Holland  Mathematics Studies 84)}, Amsterdam, North-Holland, 1984.

\bibitem{OvCo98}
J.~M. Overduin and F.~I. Cooperstock,
Phys. Rev. {\bf D58} (1998) 043506.

\bibitem{BaIs91}
C.~Barrabes and W.~Israel,
Phys. Rev. {\bf D43} (1991) 1129.

\bibitem{AdPe78}
S.~L. Adler and R.~B. Pearson,
Phys. Rev. {\bf D18} (1978) 2798.

\bibitem{Gi91}
G.~W.~Gibbons,
Lecture Notes in Phys., {\bf 383},~110-133.
Springer-Verlag, Berlin, 1991.

\bibitem{La93}
A.~Lahiri,
Mod. Phys. Lett. {\bf A8} (1993) 1549.

\bibitem{BeLe95}
O.~Bechmann and O.~Lechtenfeld,
Class. Quant. Grav. {\bf 12} (1995) 1473.

\bibitem{DeLe96}
H.~Dennhardt and O.~Lechtenfeld,
Int. J. Mod. Phys. {\bf A13} (1998) 741-764.

\bibitem{ToMaNa99}
T.~Torii, K.~Maeda and M.~Narita,
Phys. Rev. {\bf D59} (1999) 064027.

\bibitem{MaBe96}
A.~E.~Mayo and J.~D. Bekenstein,
Phys. Rev. {\bf D54} (1996) 5059.

\bibitem{Be96}
 J.~D. Bekenstein,
in: Proc. of the Second Sakharov Conference in Physics, Moscow, 1966.
Eds. I.~M.~Dremin and A.~M.~Semihatov, World Scientific, Singapore, 1997.
gr-qc/9605059.


\end{thebibliography}
\end{document}